\documentclass[smallextended]{svjour3}       % onecolumn (second format)
\smartqed  % flush right qed marks, e.g. at end of proof
\usepackage{graphicx}

\usepackage{amsfonts}
\usepackage{amsmath}
\usepackage{supertabular}

\newcommand{\JCPtoCPL}[4]{{#1} {#2} ({#4}) {#3}.}

\newcommand{\Ref}[4]{\JCPtoCPL{#1}{#2}{#3}{#4}}

\newcommand{\jcp}[3]{\Ref{J. Chem. Phys}{#1}{#2}{#3}}

\newcommand{\cpl}[3]{\Ref{Chem. Phys. Lett.}{#1}{#2}{#3}}

\newcommand{\jmathchem}[3]{\Ref{J. Math. Chem.}{#1}{#2}{#3}}

\newcommand{\jmolspec}[3]{\Ref{J. Mol. Spectrosc.}{#1}{#2}{#3}}

\newcommand{\ijqc}[3]{\Ref{Int. J. Quantum Chem.}{#1}{#2}{#3}}

\newcommand{\physrev}[3]{\Ref{Phys. Rev.}{#1}{#2}{#3}}

\def\a{\alpha}
\def\b{\beta}

\begin{document}

\title{ 
Spin contamination and noncollinearity in general complex Hartree-Fock wave functions
}
%\subtitle{Do you have a subtitle?\\ If so, write it here}

%\titlerunning{Short form of title}        % if too long for running head

\author{Patrick Cassam-Chena\"\i}
% \author{First Author         \and
%         Second Author %etc.
% }

%\authorrunning{Short form of author list} % if too long for running head

% %\institute{F. Author \at
%               first address \\
%               Tel.: +123-45-678910\\
%               Fax: +123-45-678910\\
%               \email{fauthor@example.com}           %  \\
% %             \emph{Present address:} of F. Author  %  if needed
%            \and
%            S. Author \at
%               second address
% }

\institute{Patrick Cassam-Chena\"\i \at Univ. Nice Sophia Antipolis, CNRS,  LJAD, UMR 7351, 06100 Nice, France\\
Tel.: +33-4-92076260\\
Fax: +33-4-93517974\\
\email{cassam@unice.fr}
}
%\address{Univ. Nice Sophia Antipolis, CNRS,  LJAD, UMR 7351, 06100 Nice, France.  cassam@unice.fr}

\date{\today}
%\date{Received: date / Accepted: date}
% The correct dates will be entered by the editor

\maketitle

\begin{abstract}
An expression for the square of the spin operator expectation value, $\langle S^2 \rangle$, is obtained for a general complex Hartree-Fock (GCHF) wave function and decomposed into four contributions: The main one whose expression is formally identical to the restricted (open-shell) Hartree-Fock expression. A spin contamination one formally analogous to that found for spin unrestricted  Hartree-Fock wave functions. A noncollinearity contribution related to the fact that the wave function is not an eigenfunction of the spin-$S_z$ operator. A perpendicularity contribution related to the fact that the spin density is not constrained to be zero in the xy-plane. All these contributions are evaluated and compared for the H$_2$O$^+$ system. The optimization of the collinearity axis is also considered.
\keywords{ spin contamination \and collinearity \and general complex Hartree-Fock}
% \PACS{PACS code1 \and PACS code2 \and more}
% \subclass{MSC code1 \and MSC code2 \and more}
\end{abstract}
%\begin{frontmatter}

%  
% 
% \begin{abstract}      % Produces section heading.  Lower-level
% An expression for the square of the spin operator expectation value, $\langle S^2 \rangle$, is obtained and analysed for a general complex Hartree-Fock
% (GCHF) wave function.
% \end{abstract}      % Produces section heading.  Lower-level
% 
% \begin{keyword}
%  spin contamination \sep general complex Hartree-Fock
% \end{keyword}

%\end{frontmatter}
\newpage
\section{Introduction}
Particle-independent models based on single Slater determinant wave functions, have enjoyed considerable
interest in quantum chemistry, since the pioneering works of Hartree, Slater and Fock \cite{Hartree28,Slater29,Fock30}. 

When a quantum system is described by a spin-free Hamiltonian, which obviously commutes with the spin operators $S_z$ and $S^2$, a spin-symmetry respectful way of using the Hartree-Fock method consists in:\\
1) Using spin-orbitals of pure $\a$- or $\b$-spin, so that the HF optimized Slater determinant is an eigenfunction of  $S_z$;\\
2) Imposing the spin-equivalence restriction \cite{Berthier64}, which means that paired $\alpha$- and $\beta$-spin-orbitals are formed from the same set of linearly independent orbitals. We have proved mathematically \cite{Cassam92,Cassam92-cras,Cassam93-ijqc} that this additional constraint is a necessary and sufficient condition to insure that a Slater determinantal wave function is an eigenfunction of the spin operator $S^2$. In other words, we have shown that relaxing the $S^2$-symmetry constraint exactly amounts to allow different ``paired orbitals'', in the sense of Refs. \cite{Amos61,Lowdin62}, to have different spins. This equivalence enabled us to characterize the variational space explored by the restricted open-shell Hartree-Fock (ROHF) method \cite{Cassam94}, which precisely consists in optimizing a Slater determinant subject to constraints 1) and 2) (plus spatial-symmetry constraints if any) \cite{Roothaan60}. The equivalence was also discovered independently \cite{Andrews91} by optimizing a Slater determinant with a Lagrange multiplier, enforcing $\langle S^2\rangle$ to be arbitrarily close to the ROHF value, instead of applying the spin-equivalence restriction. Not surprisingly, the determinant was approaching the ROHF solution.

Similar to the spin-free case is the ``complex-free'' one: When a quantum system is described by a real Hamiltonian, which obviously commutes with complex conjugation, one can restrict oneself to the calculation of real eigenfunctions. Then, it is  also possible to  employ only real spin-orbitals to construct the HF Slater determinant \cite{Brandas68}. (However, difficulties may occur when the symmetry group of the molecule cannot be represented over real numbers, and nonetheless, one wishes the  spin-orbitals to be adapted to spatial-symmetry).

However, it has been proposed by various authors to relax some or all of the  above-mentioned constraints, to gain variational freedom.  For example, the  different orbitals for different spins method (DODS) of Refs. \cite{Berthier54,Pople54}, (which is usually just called ``unrestricted Hartree-Fock'' (UHF), but in this paper we use ``DODS'' to avoid confusions), relaxes the spin-equivalence restriction, hence the HF solution is no longer an eigenfunction of $S^2$. Other authors \cite{Bunge67,Lefebvre67,Lunell72} have advocated the use of general spin-orbitals, mixing  $\a$-spin and $\b$-spin parts, in conjunction with the use of projectors \cite{Lowdin55}.  

Along the same line of thought, the use of  complex spin-orbitals has been proposed \cite{Lefebvre67,Hendekovic74}  to increase variational freedom in the case of real Hamiltonian. Prat and Lefebvre went a step further with so-called ``hypercomplex'' spin-orbitals to construct Slater determinants of arbitrary accuracy  \cite{Prat69}. However, the coefficients of their spin-orbitals were elements of a Clifford algebra of dimension $2^{2n}$, that was not a normed division algebra,  also known as Cayley algebra, for arbitrary values of $n$. This was unfortunate, since such a structure appears to be a minimal requirement for a quantum formalism, if, for example, Born's interpretation of the wave function is to hold firmly. For $n=1$, the Clifford algebra of Prat et al. was actually the non-commutative field of Quaternions, therefore, \textit{a fortiori}, a normed division algebra. The only larger normed division algebra is the Octonion algebra. It is a Clifford algebra of dimension $8$, which has also been proposed in a quantum mechanical 
context \cite{Penney68}, but this algebra is neither commutative nor associative. The lack of these properties rises difficulties for its use for multipartite quantum systems, nevertheless these difficulties can be overcome by keeping the product of octonion coefficients in the form of a tensor product. So, octonion-unrestricted HF appears to be the largest Cilfford algebra-unrestricted single determinantal method that can be considered in the spirit of Prat and Lefebvre's proposal. However, octonions seem incompatible with the desirable requirement that the algebra of quantum observables be what is now called a formally real Jordan algebra \cite{Jordan34} acting on a vector space of arbitrarily large dimension. Octonions are also ruled out by the requirement of orthomodularity in infinite dimension according to Sol\`er's theorem \cite{Soler95,Holland95}, which restricts quantum Hilbert spaces to be real, complex or at most quaternionic.

The first HF molecular calculations with general complex spin-orbitals, without projecting out the symmetry-breaking part of the wave function,  are  maybe those of Ref. \cite{Mayer93}. It was found on the BH molecule around its equilibrium geometry that the general complex Hartree-Fock (GCHF)  energy  was indeed lower than the DODS one, which itself was lower than the restricted Hartree-Fock (RHF) solution. So necessarily, the corresponding GCHF wave functions had $S^2$-spin contamination and $S_z$-spin contamination, that is to say, the expectation values of these operators were different from $0$, the value expected for a singlet ground state. (It is not clear whether complex numbers were used for this molecule, but the authors did mention that they performed complex calculations for $2$-electron systems.)

Relaxing the  ``$S_z$-constraint'' hence the ``collinearity constraint'', becomes perfectly legitimate
 when hyperfine or spin-orbit couplings are considered,  since the operator $S_z$ no longer commutes with the Hamiltonian.
 As a matter of fact, real physical systems do exhibit  either light \cite{Cassam02-jcp,Cassam12b-jcp} or strong \cite{Coey87,Libby91} noncollinearity of their spin densities.
Similarly, the use of complex spin-orbitals is natural, when considering relativistic corrections resulting in a complex Hamiltonian operator. So, in such a context, one should use no less than general complex spin-orbitals in  HF calculations \cite{Jayatilaka98}. The ``spin-same-orbit'' coupling term used in these calculations does not commute with the  $S^2$-operator. Therefore, one cannot strictly speak  of ``$S^2$-spin contamination'' in relativistic GCHF wave functions. However, calculating the expectation value of $S^2$, a \textit{bona fide} quantum observable, can still provide valuable physical information about the system.

A general expression for the expectation value of $S^2$ has been obtained in the DODS case \cite{Amos61}, and has served as a measure of $S^2$-spin contamination. However, as far as we are aware, no such formula has been published in the case of a GCHF wave function. This gap will be filled in the next section. 

Studying departure from collinearity is more difficult because of arbitrariness in the quantification axis. One possible way to overcome the difficulty would be to apply an external magnetic field to fix the $z$-axis but small enough not to perturb the GCHF solution. However, an elegant alternative has been proposed recently by Small et al. \cite{Small2015}. It is based on studying the lowest eigenvalue of a $(3\times 3)$-matrix built from expectation values of spin operator components and their products. In the GCHF case, the authors provided the expressions required to compute the matrix elements in 
%terms of the
%reduced density matrices  by using second quantization
%atomic orbital overlap matrix and molecular orbital coefficients
a compact form. In the third section, we give a more extended formula
%directly 
in terms of molecular orbital overlap matrix elements. We also illustrate the connections between spin contamination,
noncollinearity and its correlative: ``perpendicularity'' on the H$_2$O$^+$ cation example.
We sum up our conclusions in the last section.  

\section{Spin contamination in GCHF}

A General Complex Hartree Fock (GCHF) wave function

\begin{equation}
 \Phi_{GCHF}=\phi_1\wedge\cdots\wedge\phi_{N_e}
\label{GCHF-wf}
\end{equation} 
is the antisymmetrized product (or wedge product, denoted by $\wedge$) of orthonormal spinorbitals,
or ``two-component spinors'',

\begin{equation}
\phi_i=
\left( \begin{array}{c} \phi_{i \alpha} \\ \phi_{i \beta} \end{array}\right),
\label{MOi2c}
\end{equation}

\begin{equation}
\langle \phi_i|\phi_j\rangle=\delta_{i,j},
\label{orthonormal}
\end{equation} 
%\noindent
where the scalar product $\langle \cdot |\cdot\rangle$ means integration over space variables and summation (i.e. taking the trace) over spin variables: $\langle \phi_i|\phi_j\rangle=\langle \phi_{i \alpha}|\phi_{j \alpha}\rangle+\langle \phi_{i \beta}|\phi_{j \beta}\rangle$, (where the same bracket symbol is used for the scalar product between orbital parts). We define the ``number of $\alpha$-spin electrons'' (respectively ``number of  $\beta$-spin electrons'') as $N_\alpha:= \sum\limits_{i=1}^{N_e}\langle\phi_{i \alpha}|\phi_{i \alpha}\rangle$ (respectively, $N_\beta:= \sum\limits_{i=1}^{N_e}\langle\phi_{i \beta}|\phi_{i \beta}\rangle$). It is the expectation
value of the projection operator on the $\alpha$- (respectively $\beta$-) one-electron Hilbert subspace (more rigorously speaking, the operator induced onto the $n$-electron Hilbert space by this one-electron projection operator). Note that these two numbers need not be integer numbers, however their sum is an integer: $N_\alpha+N_\beta=N_e$.

Let us work out the expectation value of the spin operator,
\begin{equation}
S^2=S_z^2+\frac{1}{2}(S^+S^- + S^-S^+),
\label{S2}
\end{equation} 
on a general GCHF wave function.\\
The action of $S_z$ is given by,
\begin{equation}
S_z \Phi_{GCHF}=\frac{1}{2}\sum\limits_{i=1}^{N_e}\hat{\Phi}_{GCHF}^{i},
\label{Sz}
\end{equation} 
where,
\begin{equation}
\hat{\Phi}_{GCHF}^{i}=\phi_1\wedge\cdots\wedge\phi_{i-1}\wedge\hat{\phi}_{i}\wedge\phi_{i+1}\wedge\cdots\wedge\phi_{N_e},
\label{hat-i}
\end{equation} 
and,
\begin{equation}
\hat{\phi}_{i}=
\left( \begin{array}{c} +\phi_{i \alpha} \\ -\phi_{i \beta} \end{array}\right).
\label{MO-hat-i}
\end{equation}
Note that,
\begin{equation}
\langle \hat{\phi}_{i}|\hat{\phi}_{j}\rangle=\langle \phi_i|\phi_j\rangle=\delta_{i,j}.
\label{orthonorm-hat}
\end{equation} 
So, the expectation value of $S_z$ is 

\begin{eqnarray}
\langle\Phi_{GCHF}|S_z|\Phi_{GCHF}\rangle=\frac{1}{2}\sum\limits_{i=1}^{N_e}\langle \Phi_{GCHF}| \hat{\Phi}_{GCHF}^{i}\rangle=\frac{1}{2}\sum\limits_{i=1}^{N_e}\langle \phi_{i}| \hat{\phi}_{i}\rangle=\frac{1}{2}\sum\limits_{i=1}^{N_e}\left(\langle \phi_{i \alpha}|\phi_{i \alpha}\rangle-\langle \phi_{i \beta}|\phi_{i \beta}\rangle\right)=\frac{N_\alpha-N_\beta}{2}.\nonumber\\
\label{Sz-expect}
\end{eqnarray} 

and that of $S_z^2$: 
\begin{eqnarray}
\lefteqn{\langle\Phi_{GCHF}|S_z^2|\Phi_{GCHF}\rangle=\langle S_z\Phi_{GCHF}|S_z\Phi_{GCHF}\rangle}\nonumber\\
&=\frac{1}{4}\sum\limits_{i,j=1}^{N_e}\langle \hat{\Phi}_{GCHF}^{i}| \hat{\Phi}_{GCHF}^{j}\rangle&\nonumber\\
&=\frac{1}{4}\left(\sum\limits_{i=1}^{N_e}\langle \hat{\Phi}_{GCHF}^{i}| \hat{\Phi}_{GCHF}^{i}\rangle+\sum\limits_{\stackrel{i,j=1}{i\neq j}}^{N_e}\langle \hat{\Phi}_{GCHF}^{i}| \hat{\Phi}_{GCHF}^{j}\rangle\right)&\nonumber\\
&=\frac{1}{4}\sum\limits_{i=1}^{N_e}\left(\langle \hat{\phi}_{i}| \hat{\phi}_{i}\rangle+\sum\limits_{\stackrel{j=1}{j\neq i}}^{N_e}(-1)\arrowvert\langle \hat{\phi}_{i}| \phi_{j}\rangle\arrowvert^2+\langle \hat{\phi}_{i}| \phi_{i}\rangle\langle \phi_{j}| \hat{\phi}_{j}\rangle\right)&\nonumber\\
&=\frac{1}{4}\left(N_e+\sum\limits_{\stackrel{i,j=1}{i\neq j}}^{N_e}(-1)\arrowvert\langle\phi_{i \alpha}| \phi_{j \alpha}\rangle-\langle\phi_{i \beta}| \phi_{j \beta}\rangle\arrowvert^2+  \left(\langle\phi_{i \alpha}| \phi_{i \alpha}\rangle-\langle\phi_{i \beta}| \phi_{i \beta}\rangle\right)\left(\langle \phi_{j\alpha}| \phi_{j \alpha}\rangle-\langle \phi_{j \beta}| \phi_{j \beta}\rangle\right)\right)&\nonumber\\
&=\frac{1}{4}\left(N_e+\sum\limits_{i,j=1}^{N_e} \left(\langle\phi_{i \alpha}| \phi_{i \alpha}\rangle-\langle\phi_{i \beta}| \phi_{i \beta}\rangle\right)\left(\langle \phi_{j\alpha}| \phi_{j \alpha}\rangle-\langle \phi_{j \beta}| \phi_{j \beta}\rangle\right)-\arrowvert\langle\phi_{i \alpha}| \phi_{j \alpha}\rangle-\langle\phi_{i \beta}| \phi_{j \beta}\rangle\arrowvert^2\right) &\nonumber\\
&=\left(\frac{N_\alpha}{2}-\frac{N_\beta}{2}\right)^2+\frac{1}{4}\left(N_e-\sum\limits_{i,j=1}^{N_e} \arrowvert\langle\phi_{i \alpha}| \phi_{j \alpha}\rangle-\langle\phi_{i \beta}| \phi_{j \beta}\rangle\arrowvert^2\right) .&
\label{Sz2}
\end{eqnarray} 
This equation reduces to $\left(\frac{N_\alpha}{2}-\frac{N_\beta}{2}\right)^2$ in the case of a DODS wave function. So, the second term on the right-hand side (rhs), which is $(\langle\Phi_{GCHF}|S_z^2|\Phi_{GCHF}\rangle-\langle\Phi_{GCHF}|S_z|\Phi_{GCHF}\rangle^2)$ is directly related to relaxation of the $S_z$-constraint and will be called the ``$z$-noncollinearity'' contribution. Note, however, that for a GCHF wave function, the first term on the rhs  does not necessarily correspond to an eigenvalue of $S_z^2$, according to the definition of $N_\a$ and $N_\b$.\\
The action of $S^+$ is given by,
\begin{equation}
S^+ \Phi_{GCHF}=\sum\limits_{i=1}^{N_e}\acute{\Phi}_{GCHF}^{i},
\label{S+}
\end{equation} 
where,
\begin{equation}
\acute{\Phi}_{GCHF}^{i}=\phi_1\wedge\cdots\wedge\phi_{i-1}\wedge\acute{\phi}_{i}\wedge\phi_{i+1}\wedge\cdots\wedge\phi_{N_e},
\label{i+}
\end{equation} 
and,
\begin{equation}
\acute{\phi}_{i}=
\left( \begin{array}{c} +\phi_{i \beta} \\ 0 \end{array}\right).
\label{MO-i+}
\end{equation}
Similarly, the action of $S^-$ is given by,
\begin{equation}
S^- \Phi_{GCHF}=\sum\limits_{i=1}^{N_e}\grave{\Phi}_{GCHF}^{i},
\label{S-}
\end{equation} 
where,  
\begin{equation}
\grave{\Phi}_{GCHF}^{i}=\phi_1\wedge\cdots\wedge\phi_{i-1}\wedge\grave{\phi}_{i}\wedge\phi_{i+1}\wedge\cdots\wedge\phi_{N_e},
\label{i-}
\end{equation} 
and,
\begin{equation}
\grave{\phi}_{i}=
\left( \begin{array}{c} 0 \\ +\phi_{i \alpha} \end{array}\right).
\label{MO-i-}
\end{equation}
So, the expectation value of $S^-S^+$ is,

\begin{eqnarray}
\lefteqn{ \langle\Phi_{GCHF}|S^-S^+|\Phi_{GCHF}\rangle=\langle S^+\Phi_{GCHF}|S^+\Phi_{GCHF}\rangle }\nonumber\\
&=\sum\limits_{i,j=1}^{N_e}\langle \acute{\Phi}_{GCHF}^{i}| \acute{\Phi}_{GCHF}^{j}\rangle&\nonumber\\
&=\sum\limits_{i=1}^{N_e}\langle \acute{\Phi}_{GCHF}^{i}| \acute{\Phi}_{GCHF}^{i}\rangle + \sum\limits_{\stackrel{i,j=1}{i\neq j}}^{N_e} \langle \acute{\Phi}_{GCHF}^{i}| \acute{\Phi}_{GCHF}^{j}\rangle &\nonumber\\
&=\sum\limits_{i=1}^{N_e}\left(\langle \acute{\phi}_{i}| \acute{\phi}_{i}\rangle+\sum\limits_{\stackrel{j=1}{j\neq i}}^{N_e}(-1)\arrowvert\langle \acute{\phi}_{i}| \phi_{j}\rangle\arrowvert^2+\langle \acute{\phi}_{i}| \phi_{i}\rangle\langle \phi_{j}| \acute{\phi}_{j}\rangle\right)&\nonumber\\
&=\sum\limits_{i=1}^{N_e}\left(\langle\phi_{i \beta}| \phi_{i \beta}\rangle+\sum\limits_{\stackrel{j=1}{j\neq i}}^{N_e}(-1)\arrowvert\langle\phi_{i \beta}| \phi_{j \alpha}\rangle\arrowvert^2+  \langle\phi_{i \beta}| \phi_{i \alpha}\rangle\langle \phi_{j\alpha}| \phi_{j \beta}\rangle\right)&\nonumber\\
&=N_{\beta}+\sum\limits_{i,j=1}^{N_e} \langle\phi_{i \beta}| \phi_{i \alpha}\rangle\langle \phi_{j\alpha}| \phi_{j \beta}\rangle- \langle\phi_{i \beta}| \phi_{j \alpha}\rangle\langle\phi_{j \alpha}| \phi_{i \beta}\rangle.&
\label{S-S+}
\end{eqnarray} 

Similarly, the expectation value of $S^+S^-$ is, 
\begin{eqnarray}
\lefteqn{ \langle\Phi_{GCHF}|S^+S^-|\Phi_{GCHF}\rangle=\langle S^-\Phi_{GCHF}|S^-\Phi_{GCHF}\rangle }\nonumber\\
&=\sum\limits_{i=1}^{N_e}\left(\langle \grave{\phi}_{i}| \grave{\phi}_{i}\rangle+\sum\limits_{\stackrel{j=1}{j\neq i}}^{N_e}(-1)\arrowvert\langle \grave{\phi}_{i}| \phi_{j}\rangle\arrowvert^2+\langle \grave{\phi}_{i}| \phi_{i}\rangle\langle \phi_{j}| \grave{\phi}_{j}\rangle\right)&\nonumber\\
&=\sum\limits_{i=1}^{N_e}\left(\langle\phi_{i \alpha}| \phi_{i \alpha}\rangle+\sum\limits_{\stackrel{j=1}{j\neq i}}^{N_e}(-1)\arrowvert\langle\phi_{i \alpha}| \phi_{j \beta}\rangle\arrowvert^2+  \langle\phi_{i \alpha}| \phi_{i \beta}\rangle\langle \phi_{j\beta}| \phi_{j \alpha}\rangle\right)&\nonumber\\
&=N_{\alpha}+\sum\limits_{i,j=1}^{N_e} \langle\phi_{i \alpha}| \phi_{i \beta}\rangle\langle \phi_{j\beta}| \phi_{j \alpha}\rangle- \langle\phi_{i \alpha}| \phi_{j \beta}\rangle\langle\phi_{j \beta}| \phi_{i \alpha}\rangle.&
\label{S+S-}
\end{eqnarray} 
Using Eq.(\ref{S2}) and putting together Eqs.(\ref{Sz2}), (\ref{S-S+}) and (\ref{S+S-}), one obtains the expectation value of $S^2$,
\begin{eqnarray}
\lefteqn{ \langle\Phi_{GCHF}|S^2|\Phi_{GCHF}\rangle=\left(\frac{N_\alpha}{2}-\frac{N_\beta}{2}\right)^2+\frac{N_\alpha}{2}+\frac{N_\beta}{2}+\frac{1}{4}\left(N_e-\sum\limits_{i,j=1}^{N_e} \arrowvert\langle\phi_{i \alpha}| \phi_{j \alpha}\rangle-\langle\phi_{i \beta}| \phi_{j \beta}\rangle\arrowvert^2\right)}\nonumber\\
&+\sum\limits_{i,j=1}^{N_e} \langle\phi_{i \alpha}| \phi_{i \beta}\rangle\langle \phi_{j\beta}| \phi_{j \alpha}\rangle- \langle\phi_{i \alpha}| \phi_{j \beta}\rangle\langle\phi_{j \beta}| \phi_{i \alpha}\rangle.&
\label{expect-S2}
\end{eqnarray} 
The expression reduces to the known formula in the case of a DODS wave function. Assuming, without loss of generality, that  $N_\alpha\geq N_\beta$, we rewrite Eq. (\ref{expect-S2}) as,
\begin{eqnarray}
\lefteqn{ \langle\Phi_{GCHF}|S^2|\Phi_{GCHF}\rangle=\left(\frac{N_\alpha}{2}-\frac{N_\beta}{2}\right)\left(\frac{N_\alpha}{2}-\frac{N_\beta}{2}+1\right)+\frac{1}{4}\left(N_e-\sum\limits_{i,j=1}^{N_e} \arrowvert\langle\phi_{i \alpha}| \phi_{j \alpha}\rangle-\langle\phi_{i \beta}| \phi_{j \beta}\rangle\arrowvert^2\right)}\nonumber\\
&+\left(N_\beta-\sum\limits_{i,j=1}^{N_e} \langle\phi_{i \alpha}| \phi_{j \beta}\rangle\langle\phi_{j \beta}| \phi_{i \alpha}\rangle\right) + \lvert\sum\limits_{i=1}^{N_e}\langle \phi_{i\beta}| \phi_{i\alpha}\rangle\rvert^2.&
\label{spin-contam}
\end{eqnarray} 
In this formula we identify four contributions: The first term is formally identical to the ROHF expression also found in the DODS case. However, care must be taken that it is actually different, because the numbers of $\alpha$- and $\beta$-electrons are not good quantum numbers in the GCHF case.  The second term  on the first line is the ``$z$-noncollinearity'' contribution. The third term on the second line is formally analogous to the ``spin contamination'' of a DODS wave function as defined in \cite{Cassam93-ijqc,Amos61}. Finally, the last term on the second line is the square of the expectation value of the lowering or raising operator:
\begin{eqnarray}
\lvert\sum\limits_{i=1}^{N_e}\langle \phi_{i\beta}| \phi_{i\alpha}\rangle\rvert^2=\lvert\langle\Phi_{GCHF}|S^+|\Phi_{GCHF}\rangle\rvert^2=\lvert\langle\Phi_{GCHF}|S^-|\Phi_{GCHF}\rangle\rvert^2.
\label{nonperp}
\end{eqnarray}
A non zero contribution of this term can only arise from the release of the $S_z$-constraint, which allows for the $\alpha$- and $\beta$-components of a given, general spin-orbital to be both non zero. But it originates from $S^+S^-$ and $S^-S^+$, and is maximal when $\phi_{i\beta}=\exp(\imath\theta)\phi_{i\alpha}$ for all i, that is to say  when the $\phi_i$'s are eigenfunctions of $cos\theta S_x + sin\theta S_y$ for some angle $\theta$. It is related to the emergence of a non-zero spin density in the $x,y$-plane, correlatively to the loss of $z$-collinearity. We tentatively call this term the ``$x,y$-perpendicularity'' contribution.

% This formula shows that, compared to the DODS spin contamination formula \cite{Amos61}, besides the ``$z$-noncollinearity'' contribution  on the first line, there is an extra term (middle term on the second line) arising from the possible non-orthogonality of the two components of a given, general spin-orbital. However, care must be taken that the first term, which is formally identical to the DODS case, is actually different, since the numbers of $\alpha$- and $\beta$-electrons are not good quantum numbers in the GCHF case.

The present formulas have been implemented in the code TONTO \cite{Tonto} and applied in a recent article (third column of Tab. 4 in \cite{Bucinsky15}).
Let us discuss further the different contributions to $S^2$ for a  $H_2O^+$ GCHF calculation similar to that reported in \cite{Bucinsky15}.
The $z$-quantification axis was the axis perpendicular to the plane of the molecule.  The results, see Tab. \ref{tab-contam}, shows that the main contribution to $\langle\Phi_{GCHF}|S^2|\Phi_{GCHF}\rangle$ beside the reference expression (first term in eq.(\ref{spin-contam})) is the so-called spin contamination contribution (we set $\hbar=1$ throughout the paper). The $x,y$-perpendicularity and $z$-noncollinearity contributions are of the same order of magnitude and more than one order of magnitude smaller. Added to $\left(\frac{N_\alpha}{2}-\frac{N_\beta}{2}\right)\left(\frac{N_\alpha}{2}-\frac{N_\beta}{2}+1\right)$, they almost make up the reference value of $+0.75$. So, the spin contamination value of $0.007033$ amounts almost exactly to the difference between the exact expectation value $\langle\Phi_{GCHF}|S^2|\Phi_{GCHF}\rangle$ and this reference ``ROHF value''. This demonstrates that, in Table 4 of Ref.\cite{Bucinsky15}, the equality of the entries in column 2 (reference ROHF value plus spin contamination term) and column 4 (our $\langle\Phi_{GCHF}|S^2|\Phi_{GCHF}\rangle$ value) does not imply no noncollinearity. In contrast, if for a given line of the table, these two quantities differ, then necessarily there will be some noncollinearity  in the corresponding GCHF wave function. This can be shown \textit{reductio ad absurdum}. Suppose that the $z$-collinearity constraint is fullfiled, then $N_\alpha$ and $N_\beta$ will be good quantum numbers and the first term in our expression of $\langle\Phi_{GCHF}|S^2|\Phi_{GCHF}\rangle$ will be equal to the ROHF reference value. The spin contamination contribution being included in both quantities, the difference between them must arise from the $x,y$-perpendicularity and $z$-noncollinearity contributions. One at least of the contributions arising from the release of the collinearity constraint must be non zero, hence a contradiction. This hints that the following systems $Cl$, $HCl^+$, $Fe$, $Cu$, $Cu^{2+}$ and $[OsCl_5(Hpz)]^-$ reported in Table 4 of Ref.\cite{Bucinsky15} would present stronger non collinearity than $H_2O^+$.

\section{Collinearity in GCHF}

In the previous section, we have encountered a $z$-(non)collinearity measure, $col_z:=(\langle S_z^2\rangle-\langle S_z \rangle^2)$. This quantity can be generalized to an arbitrary quantization direction defined by a unit vector $\vec{u}=\left( \begin{array}{c} u_x \\ u_y\\ u_z \end{array}\right) $ of the unit sphere $\mathcal{S}^2$ of $\mathbb{R}^3$ by replacing $S_z$ by $\vec{u}\cdot \vec{S}=\sum\limits_{\mu\in\{x,y,z\}}u_\mu  S_\mu$. Then, $\vec{u}$-(non)collinearity  is measured by: 
\begin{equation}
col(\vec{u}):=\sum\limits_{\mu,\nu\in\{x,y,z\}}u_\mu u_\nu(\langle S_\mu S_\nu\rangle-\langle S_\mu \rangle\langle S_\nu \rangle). 
\end{equation}
Small et al. \cite{Small2015} defined a (non)collinearity measure by:
\begin{equation}
col:=\min\limits_{\vec{u}\in\mathcal{S}^2}\ col(\vec{u}), 
\end{equation}
which corresponds to the lowest eigenvalue of the matrix $A$ whose elements are given by,
\begin{equation}
A_{\mu\nu}=\Re(\langle S_\mu S_\nu\rangle)-\langle S_\mu \rangle\langle S_\nu \rangle,
\end{equation}
where $\Re(z)$ is the real part of $z$. The associated eigenvector gives the optimal collinearity direction. Setting
%$^\mu\tilde{\phi}=S_\mu\phi$, that is to say 
$^x\tilde{\phi}=\frac{1}{2}(\acute{\phi}+\grave{\phi})$,  $^y\tilde{\phi}=\frac{-\imath}{2}(\acute{\phi}-\grave{\phi})$ and  $^z\tilde{\phi}=\hat{\phi}$, we have in this notation,
\begin{equation}
\forall \mu, \nu \in \{x,y,z\}\qquad A_{\mu\nu}=\delta_{\mu\nu}\frac{N_e}{4}-\sum\limits_{i,j=1}^{N_e} \langle^\mu\tilde{\phi}_i| \phi_{j}\rangle\langle\phi_{j}| ^\nu\tilde{\phi}_i\rangle,
\end{equation}
where $\delta_{\mu\nu}$ is the Kr\"onecker symbol.\\

\noindent
Returning  to the  $H_2O^+$ example and applying these formulae, we obtain 
\begin{equation}
A=\begin {pmatrix}
+0.253128      &    +0.000145     &     -0.009774   \\
+0.000145      &    +0.253451     &     +0.003745   \\
-0.009774      &    +0.003745     &     +0.000461   
\end{pmatrix}
\end{equation}
The diagonalization of the $A$-matrix gives the optimal collinear direction:
\begin{equation} 
\vec{u_0}^t=\left(+0.0385908, -0.014789,+0.999146 \right) ,
\end{equation}
which is only slightly tilted with respect to the $z$-direction, and the system is quasi-collinear in this direction since $col=0.000028$ is very close to zero. This shows that the noncollinearity contribution to $\langle\Phi_{GCHF}|S^2|\Phi_{GCHF}\rangle$ could be further reduced by more than one order of magnitude by selecting the optimal quantization axis corresponding to $\vec{u_0}$ instead of the spatial $z$-axis. The perpendicularity contribution would decrease accordingly.

\section{Conclusion}

We have decomposed the expectation value of the spin operator $S^2$ into (i) a term formally identical to its expression for a ROHF
reference wave function, (ii) a term called ``spin contamination'' because it is formally analogous to that derived by Amos and Hall \cite{Amos61} for DODS wave functions, (iii) a noncollinear contribution which can be minimized by following a procedure recently introduced \cite{Small2015},  (iv) a term called the ``perpendicularity contribution'' which arises from the release of the $z$-collinearity constraint but which should rather be regarded as arising from the release of the ``nonperpendicularity constraint'' on the spin-density.  The collinearity and nonperpendicularity constraints are correlatives.

We have evaluated these four different contributions for a GCHF calculation on the $H_2O^+$ cation. Note that we used the IOTC relativistic Hamiltonian  \cite{Barysz01} so that the term ``spin contamination'' is not really appropriate in this context,  departure from the ROHF reference value being legitimate. However, the so-called spin contamination contribution has been found to dominate the noncollinearity and perpendicularity ones. This could be made even more so, by tilting the quantification axis to the optimal collinearity direction.

\section*{Acknowledgements}

We acknowledge Dr. Luk\'a\v{s} Bu\v{c}insk\'y for drawing our attention to the problem of the derivation of GCHF spin contamination, and to the fact that $S^2$ does not commutes with the ``spin-same-orbit'' coupling term, usually used in quantum chemistry.
The referees and the editor are 
acknowledged for suggesting many improvements to the manuscript.
This article is a tribute to Prof. P. Surj{\`a}n and is also dedicated to the memory of the late Prof. Gaston Berthier, 
who introduced to the author in the course of vivid discussions, many of the references listed below.

%\section*{TABLES}
\begin{table}[ht]

\begin{center}

% \begin{tabular}{cccc}
% % %\begin{xtab}{cccc}
% % \multicolumn{2}{c}{ \large{\textbf{Convergence of ground state energy}}}\\
% \hline
% \multicolumn{2}{c}{ \textbf{real}}&\multicolumn{2}{c}{ \textbf{imaginary}}\\
% \hline  
% $O$     &$\langle O \rangle$ & $O$ & $\langle O \rangle$         \\
% $N_\alpha$   & $+4.99954553$ &          &                                 \\
% $N_\beta$   & $+4.00045447$ &          &                                 \\
% %$(\frac{N_\alpha}{2}-\frac{N_\beta}{2})^2$   & 0.24954573 &          &                                 \\
% $(\frac{N_\alpha}{2}-\frac{N_\beta}{2})(\frac{N_\alpha}{2}-\frac{N_\beta}{2}+1)$   & $+0.74909126$ &          &      \\
% $z$-noncollinearity& $+0.00046085$             &&                                \\
% $\parallel$-spin contamination& $+0.00703349$             &&                                \\
% $\perp$-spin contamination& $+0.00042695$             &&                                \\
% $S^2$   &$+0.75701255$ &          &                                 \\
% $S_x^2$ &$+0.25350071$ & $S_xS_y$ & $-0.00014267 +0.24977276\ \imath$ \\
% $S_y^2$ &$+0.25350525$ & $S_yS_x$ & $-0.00014267 -0.24977276\ \imath$ \\      
% $S_z^2$ &$+0.25000659$ & $S_xS_z$ & $+0.00963844 +0.00369708\ \imath$ \\     
% $S_x$   &$+0.01929442$ & $S_zS_x$ & $+0.00963844 -0.00369708\ \imath$ \\   
% $S_y$   &$-0.00739415$ & $S_yS_z$ & $-0.00369372 +0.00964721\ \imath$ \\
% $S_z$   &$+0.49954553$ & $S_zS_y$ & $-0.00369372 -0.00964721\ \imath$ \\     
% \hline
%             
% \end{tabular}

\begin{tabular}{cc}
% \hline
% $O$     &$\langle O \rangle$         \\
\hline  
$N_\alpha$   & $+4.999546$   \\
$N_\beta$   & $+4.000454$ \\
$(\frac{N_\alpha}{2}-\frac{N_\beta}{2})(\frac{N_\alpha}{2}-\frac{N_\beta}{2}+1)$   & $+0.749091$  \\
$z$-noncollinearity& $+0.000461$    \\
$x,y$-nonperpendicularity& $+0.000427$  \\
spin contamination& $+0.007033$   \\
$\langle S^2\rangle$   &$+0.757013$  \\
% $S_x^2$ &$+0.25350071$ & $S_xS_y$ & $-0.00014267 +0.24977276\ \imath$ \\
% $S_y^2$ &$+0.25350525$ & $S_yS_x$ & $-0.00014267 -0.24977276\ \imath$ \\      
% $S_z^2$ &$+0.25000659$ & $S_xS_z$ & $+0.00963844 +0.00369708\ \imath$ \\     
% $S_x$   &$+0.01929442$ & $S_zS_x$ & $+0.00963844 -0.00369708\ \imath$ \\   
% $S_y$   &$-0.00739415$ & $S_yS_z$ & $-0.00369372 +0.00964721\ \imath$ \\
% $S_z$   &$+0.49954553$ & $S_zS_y$ & $-0.00369372 -0.00964721\ \imath$ \\     
\hline          
\end{tabular}
\end{center}
\bigskip

\caption{Expectation value of $\langle S^2\rangle$ and related quantities for an H$_2$O$^+$ GCHF optimized wave function.
The geometry parameters were $r_{OH}= 0.99192 \mathring{A}$, $\widehat{HOH}=101.411 deg$. The basis set consisted of the primitives Gaussian functions left uncontracted of Dunning's cc-pVDZ hydrogen and oxygen basis sets \cite{Dunning89}. The infinite-order two-component (IOTC) relativistic Hamiltonian of Barysz and Sadlej \cite{Barysz01} was employed. }
\label{tab-contam}

\end{table}

\end{document}